\documentclass[final,5p,times,twocolumn]{elsarticle}
\usepackage{eurosym}
\usepackage{graphicx}
\usepackage{amssymb,bm,amsmath}
\usepackage{indentfirst}
\usepackage{epsfig}
\usepackage{epstopdf}
\usepackage{exscale}
\usepackage{relsize}
\usepackage{siunitx}
\usepackage[colorlinks=true,citecolor=black]{hyperref}
\usepackage{caption2}
\usepackage{graphicx,subfigure,amssymb,dcolumn,bm,amsmath}

\journal{Chaos, Solitons \& Fractals}
\biboptions{sort&compress}
\biboptions{}

\begin{document}

\begin{frontmatter}

\title{Robust quantum-droplet necklace clusters in three dimensions}

\author{Liangwei Dong \corref{cor1}}
\cortext[cor1]{Corresponding author.}
\ead{dlw0@163.com}
\address{Department of Physics, Zhejiang University of Science and Technology, Hangzhou,  310023, China}

\author{Dongshuai Liu}
\address{School of Physics and Electronic Engineering, Fuyang Normal University, Fuyang, 236037, China }
\address{Key Laboratory of Functional Materials and Devices for Informatics of Anhui Educational Institutions, Fuyang Normal University, Fuyang, 236037, China}
\author{Boris A. Malomed}
\address{Instituto de Alta Investigacion, Universidad de Tarapaca, Casilla 7D, Arica, Chile}

\date{\today}




\begin{abstract}
We report the existence of quasi-stable ring-shaped (necklace-shaped) clusters built, in the free space, of 3D quantum droplets (QDs) in a binary Bose-Einstein condensate, modeled by the Gross-Pitaevskii equations with the Lee-Huang-Yang corrections. The QD clusters exhibit diverse dynamical behaviors, including contraction, oscillations, and expansion, depending on the cluster's initial radius. A phase shift between adjacent QDs imparts net angular momentum to the cluster, inducing its permanent rotation. Through the energy-minimization analysis, we predict equilibrium values of the necklace radius that support persistent rotation with negligible radial pulsations. In this regime, the clusters evolve as robust entities, maintaining the azimuthal symmetry in the course of the evolution, even in the presence of considerable perturbations. Necklace ``supervortex" clusters, composed of QDs with inner vorticity $1$ and global vorticity $M$, imprinted onto the cluster, may also persist for a long time. The reported findings may facilitate the experimental realization of complex self-sustained quantum states in the 3D free space.
	
\end{abstract}


\end{frontmatter}

\section{Introduction}

The creation and stabilization of self-bound complex structures in nonlinear systems represent a fundamental challenge in many areas of physics, from
condensed matter to optics and nuclei. Among these structures, solitons, which maintain their shape through the stable balance between the
nonlinearity and diffraction, have been extensively investigated \cite{kivshar2003optical, Book2, RevModPhys.88.035002,book-2022}. A particularly
challenging objective is the production of soliton necklace clusters, built of multiple individual solitons arranged in a ring-shaped configuration, in
the two- or three-dimensional (2D or 3D) space. While the stability of a single soliton may be maintained by the straightforward balance, the
clusters imply a delicate interplay between attractive and repulsive interactions. The attraction between adjacent solitons can readily lead to
the modulational instability of the cluster, while the dominant repulsion tends to cause its disintegration.

To date, clusters of optical soliton have been predicted and experimentally observed in a variety of nonlinear settings, each providing a specific
mechanism to suppress the inherent instability \cite{soljacic1998, PhysRevLett.88.053901, PhysRevLett.89.273902, PhysRevE.67.046610, Li:20,
ZENG2020168284, PhysRevA.102.033523}. In nonlocal media, long-range interactions between solitons can induce effective repulsion, enabling the
stabilization of necklace-like chains \cite{Song:18, Zhong_2008}. In systems featuring saturable nonlinearities (e.g., photorefractive crystals) or
parity-time symmetry, appropriate potential landscapes or properly engineered gain-loss profiles can trap several soliton filaments in 2D
configurations, including squares, hexagons, and rings \cite{PhysRevLett.87.033901, PhysRevLett.94.113902}.

Parallel to optics, the advent of experiments with Bose-Einstein condensates (BECs) in ultracold gases had opened avenues for exploring nonlinear
matter-wave dynamics, with mean-field (MF) effects accurately modeled by the Gross-Pitaevskii equations (GPEs). They are similar to the nonlinear Schr\"{o}dinger equations in optics. A paradigm shift occurred with the prediction of quantum droplets (QDs), i.e., a novel class of self-bound
states stabilized in binary BECs by the Lee-Huang-Yang (LHY) corrections, which represent a dynamical contribution of quantum fluctuations around the
MF states \cite{PhysRev.106.1135, PhysRevLett.115.155302}. The LHY repulsive terms balance the MF attraction in 3D and 2D setups, preventing the collapse
and enabling the formation of stable QDs, even in the free space, as corroborated by the experiment \cite{PhysRevLett.117.100401, Science2018,
Science2018_2, PhysRevLett.120.235301}. Extensive theoretical studies of QDs have been conducted in the framework of 1D, 2D, and 3D LHY-amended GPEs,
with and without external potentials \cite{PhysRevResearch.2.033522, PhysRevLett.123.133901, ZHAO2022112481, zengjianhua, XU2022112665,Zibin2026}, see
Refs. \cite{book-2022} and \cite{Frontiers} for a review. The studies were further expanded from single QDs to multi-droplet arrays \cite{PhysRevLett.122.193902, PhysRevLett.123.160405}. Notably, quasi-stable 2D QD clusters were predicted in binary BECs, exhibiting remarkably long
lifetimes \cite{PhysRevLett.122.193902}.

Although soliton clusters may resemble higher-order multipole solitons, the two types of the nonlinear modes are fundamentally different. Soliton
clusters are constructed by arranging multiple fundamental (or vortex) solitons along a ring \cite{soljacic1998, PhysRevLett.88.053901}, whereas
multipole solitons emerge as nonlinear modes bifurcating from linear eigenstates of the system including an external potential \cite{dong2023multipole, DONG2024115499}. Moreover, while the phase difference between adjacent constituents in soliton clusters may take arbitrary values,
the multipole modes are constrained to the phase difference of $\pi$ between the poles, hence the number of poles must be even, while the
clusters admit odd numbers of the constituents.

Although soliton clusters have been predicted in optical media with the Kerr, cubic-quintic, and saturable nonlinearities in both 2D \cite{soljacic1998, PhysRevLett.88.053901,HS} and 3D \cite{Mihalache_2004} settings, quasi-stable necklace-shaped QD clusters in BECs remain largely
unexplored. To date, the theoretical study has reported a particular species of long-lived 2D QD clusters in the binary BEC \cite{PhysRevLett.122.193902}. A challenging issue is to construct robust QD clusters in the free 3D space. In this work, we show that the dynamics of 3D QD clusters can be
accurately predicted via an energy-minimization analysis. We systematically investigate clusters built of multiple fundamental or vortex QDs uniformly
placed on a ring, with equilibrium values of its radius predicted by the energy minimization. Direct simulations corroborate that necklace clusters
with these equilibrium radii, although they cannot be true stationary states, are very robust patterns, surviving for a long time. On
the other hand, if the QDs are initially placed on a ring whose radius deviates from the equilibrium value, the cluster demonstrates contraction,
oscillations, or expansion, along with permanent rotation. Thus, our study aims to bridge the gap between the established dynamics of 2D QD clusters
and the previously unexplored domain of 3D self-organized quantum-matter states, suggesting experimental realization of these complex structures.

The analysis is based on the model formulated in Section 2, which also briefly recapitulates the basic results for individual QDs. The main results
for the necklace clusters are reported in Section 3. The paper is completed by Section 4.

\section{The model}

\label{Sec2} The dynamics of 3D QDs is governed by the system of coupled LHY-amended GPEs for the two-component wave function $\Psi _{1,2}$. In
addition to the usual MF cubic nonlinearities, the GPEs incorporate quartic self-repulsion terms that represent the LHY corrections. Here, we assume
equal scattering lengths for atoms in both components of the mixture, i.e., $a_{1}=a_{2}\equiv a$. In the scaled form (with the atomic mass, Planck's
constant, and the strength of the MF cubic self-repulsion set to be $1$), the coupled equations are written as \cite{PhysRevLett.115.155302,
PhysRevLett.117.100401}
\begin{equation}
\begin{aligned} i\frac{\partial \Psi _{1}}{\partial t}=-\frac{1}{2}\nabla
^{2}\Psi _{1}+\left( |\Psi _{1}|^{2}+g_L|\Psi _{1}|^{3}\right) \Psi
_{1}-g|\Psi _{2}|^{2}\Psi _{1}, \\ i\frac{\partial \Psi _{2}}{\partial
t}=-\frac{1}{2}\nabla ^{2}\Psi _{2}+\left( |\Psi _{2}|^{2}+g_L|\Psi
_{2}|^{3}\right) \Psi _{2}-g|\Psi _{1}|^{2}\Psi _{2}, \end{aligned}
\label{Eq1}
\end{equation}%
where $\nabla ^{2}=\partial ^{2}/\partial x^{2}+\partial ^{2}/\partial y^{2}+\partial ^{2}/\partial z^{2}$, while constants $g>0$ and $g_{L}=(128/3)%
\sqrt{2/\pi }a^{3/2}$ denote the relative strengths of the MF intercomponent attraction and LHY self-repulsion, respectively. Following Ref.~\cite%
{PhysRevA.98.013612}, we select for the scaled MF cross-attraction and LHY-correction strengths physically relevant values, $g=1.75$ and $g_{%
\mathrm{L}}=0.5$, respectively. Equations (\ref{Eq1}) conserve two norms $N_{1,2}$, the total energy $E$, total linear momentum, and total angular
momentum $\mathbf{J}$, whose $z$-component is also written below:
\begin{equation}
\begin{aligned} &N_{1,2}=\iiint |\Psi_{1,2}|^{2}dxdydz, \\
&E=\frac{1}{2}\iiint [ |\nabla \Psi_1 |^{2}+|\nabla \Psi_2 |^{2} +|\Psi_1
|^{4}+|\Psi_2|^{4}+ \\ &~~~~~~~~ \frac{4}{5}g_L(|\Psi_1 |^{5}+|\Psi_2
|^{5})-2g|\Psi_1|^2|\Psi_2|^2 ] dxdydz, \\ &\mathbf{J}_{z}=i
\mathlarger{\sum}_{n=1,2}\iiint \left( y\frac{\partial \Psi_n}{\partial
x}-x\frac{\partial \Psi_n}{\partial y}\right) dxdydz, \end{aligned}
\label{Eq2}
\end{equation}

As is common in the studies of binary BECs \cite{PhysRevA.98.013612,PhysRevResearch.2.033522}, we consider the symmetric
case with $\Psi _{1}=\Psi _{2}\equiv \Psi $ and $N_{1}=N_{2}\equiv N$, for which the coupled system (\ref{Eq1}) reduces to a single equation:
\begin{equation}
i\frac{\partial \Psi }{\partial t}=-\frac{1}{2}\nabla ^{2}\Psi +\left(
1-g\right) |\Psi |^{2}\Psi +g_{L}|\Psi |^{3}\Psi .  \label{Eq3}
\end{equation}%
Stationary solutions of Eq.~(\ref{Eq3}) with real chemical potential $\mu$ are sought for as
\begin{equation}
\Psi (x,y,z,t)=\psi (x,y,z)\exp (-i\mu t),  \label{Psi}
\end{equation}%
with the complex stationary wave function $\psi (x,y,z)$ satisfying equation
\begin{equation}
\frac{1}{2}\nabla ^{2}\psi +\mu \psi +(g-1)|\psi |^{2}\psi -g_{L}|\psi
|^{3}\psi =0,  \label{Eq4}
\end{equation}%
which can be solved by means of the numerical Newton-conjugate-gradient method \cite{Book2}.

For vortex states characterized by an integer winding number (topological charge) $m$, it is natural to adopt cylindrical coordinates $\left(
r,z,\theta \right) $. Then, the substitution of $\psi (x,y,z)=w(r,z)\exp(im\theta )$ in Eq.~(\ref{Eq4}) leads to the following equation for real $%
w(r,z)$ \cite{PhysRevA.98.013612}:
\begin{equation}
\frac{1}{2}\left( \frac{\partial ^{2}}{\partial r^{2}}+\frac{1}{r}\frac{%
\partial }{\partial r}-\frac{m^{2}}{r^{2}}+\frac{\partial ^{2}}{\partial
z^{2}}\right) w+\mu w+(g-1)w^{3}-g_{L}w^{4}=0.  \label{Eq5}
\end{equation}%
Stability of the vortex states was explored by considering weakly perturbed solutions, $\Psi =[w+u\exp (\lambda t+ik\theta )+v^{\ast }\exp (\lambda
^{\ast }t-ik\theta )]\exp (im\theta -i\mu t)$, where $u$ and $v$ are eigenmodes of the small perturbation, eigenvalue $\lambda \equiv \lambda
_{r}+i\lambda _{i}$ represents the complex instability growth rate, and integer $k$ is an azimuthal perturbation index. Substituting this in Eq. (%
\ref{Eq1}) leads to the corresponding system of the linearized Bogoliubov--de Gennes equations:
\begin{equation}
\begin{aligned} &i\lambda u=-\frac{1}{2}\left[\frac{\partial^2 }{\partial
r^2}+\frac{1}{r}\frac{\partial}{\partial
r}-\frac{(m+k)^2}{r^2}+\frac{\partial^2}{\partial z^2}\right]u-\mu u \\ &
+\left(2 w^2+\frac{5}{2}g_L w^3-g w^2\right) u+\left( w^2+\frac{3}{2}g_L
w^3\right)v-g w^2(u+v), \\ & i\lambda v=+\frac{1}{2}\left[\frac{\partial^2
}{\partial r^2}+\frac{1}{r}\frac{\partial}{\partial
r}-\frac{(m-k)^2}{r^2}+\frac{\partial^2}{\partial z^2}\right]v+\mu v \\ &
-\left(2 w^2+\frac{5}{2}g_L w^3-g w^2\right) v-\left( w^2+\frac{3}{2}g_L
w^3\right)u+g w^2(u+v). \end{aligned}  \label{Eq6}
\end{equation}%
Equations (\ref{Eq6}) can be solved numerically by means of the Fourier collocation algorithm \cite{Book2}. The QD vortices are stable if all
eigenvalues $\lambda $, produced by Eq.~(\ref{Eq6}), are pure imaginary.

\begin{figure}[tbph]
\centering
\includegraphics[width=0.42\textwidth]{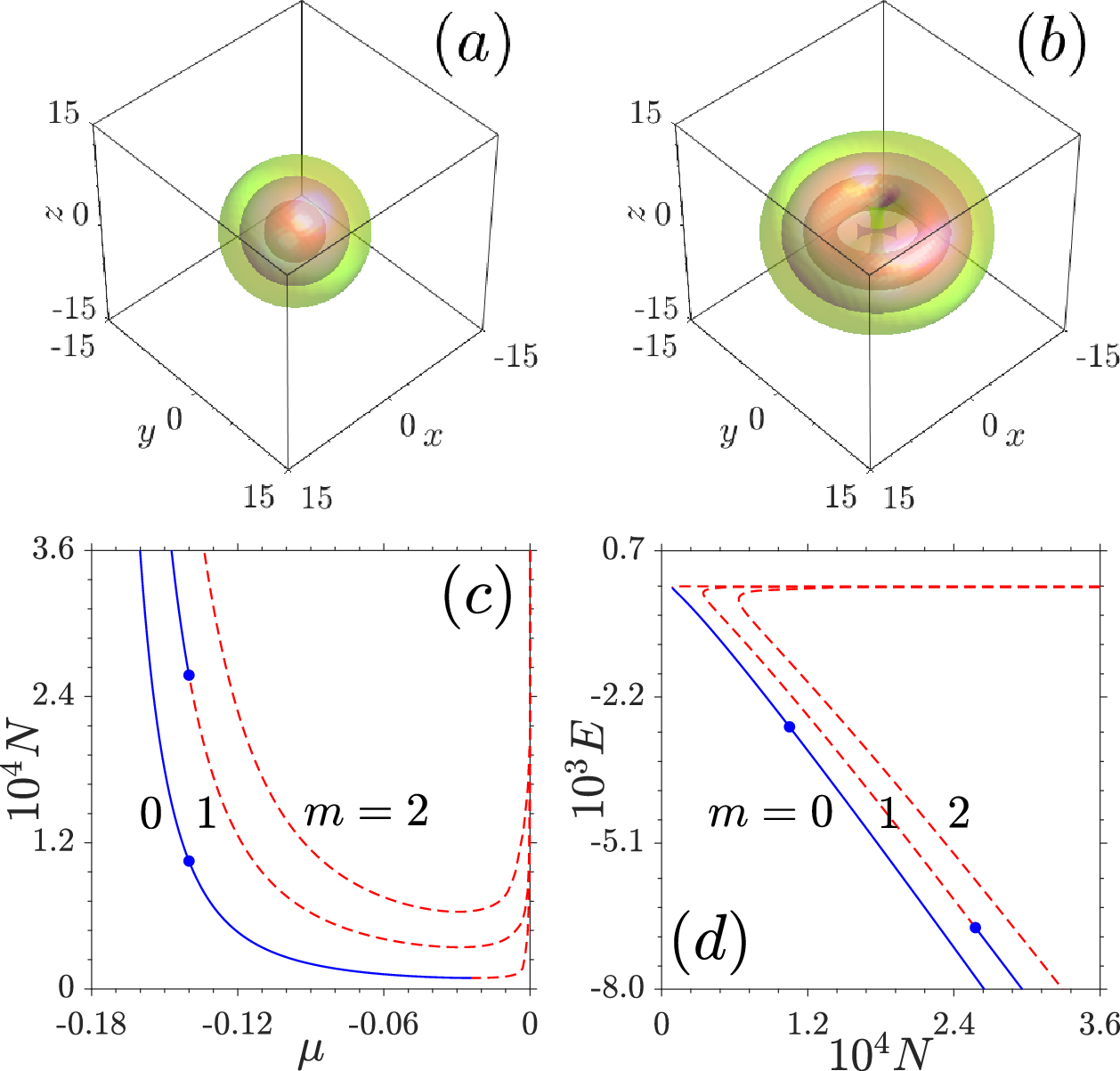}\vskip-0pc
\caption{(a) and (b): Isosurface plots of $|\protect\psi |=0.15|\protect\psi|_{\mathrm{max}}$, $0.55|\protect\psi |_{\mathrm{max}}$, and $0.95|\protect%
\psi |_{\mathrm{max}}$ for the QDs with $m=0$ and $m=1$, respectively, and $\protect\mu =-0.14$, which correspond to the points marked in panels (c) and
(d). The viewing angles are $135^{\circ }$ in azimuth and $45^{\circ }$ in elevation. (c) Norm $N$ vs. chemical potential $\protect\mu $ and (d) energy
$E$ vs. $N$ for QDs with different vorticities $m$. Solid and dashed curves represent stable and unstable branches, respectively. }
\label{Fig1}
\end{figure}

Figures \ref{Fig1}(a) and (b) display representative examples of stable QDs with topological charges $m=0$ and $1$, respectively, produced by
the numerical solution of Eq.~(\ref{Eq5}). While the one with $m=0$ naturally exhibits the spherical symmetry in Fig.~\ref{Fig1}(a), the QD with
embedded vorticity $m=1$ keeps the axial symmetry due to its topological structure, see Fig.~\ref{Fig1}(b). The norm of QDs with different
topological charges diverges as the chemical potential $\mu $ approaches the respective cutoff value, $\mu _{\text{cut}}=-(25/216)(g-1)^{3}/g_{L}^{2}%
\approx -0.1953$ \cite{PhysRevA.98.013612}, at which the QD with the flat-top shape carries over into the delocalized CW (continuous-wave) state
with the constant amplitude, see Fig.~\ref{Fig1}(c). As $\mu $ increases, the norm first decreases, attaining a minimum at a specific value of $\mu $,
and then increases again, diverging at $\mu \rightarrow 0$. Thus, QDs in the free space are purely nonlinear entities, unlike those in externally
confined systems, where both 2D and 3D QDs bifurcate from the corresponding linear eigenstates \cite{DONG2024114472, DONG2024115499}.

As shown in Fig.~\ref{Fig1}(c), the linear stability analysis, based on Eqs.~(\ref{Eq6}), shows that QDs with $m=0$ are stable when $dN/d\mu <0$, in
agreement with the well-known Vakhitov-Kolokolov (VK) criterion \cite{VK1973,Berge}. However, the VK criterion is not sufficient to ensure the
stability of the vortex QDs with $m\neq 0$ against spontaneous splitting of the vortex torus into fragments \cite{PhysRevA.98.013612}. Specifically,
while the QDs with $m=1$ are stable at $\mu \leq -0.14$, a very narrow stability interval of the vortex QDs with $m=2$ in terms of $\mu $ appears
near $\mu _{\text{cut}}$ [it is invisible on the respective curve in Fig. \ref{Fig1}(c)]. The QD energy as a function of the norm exhibits a
characteristic cusp-like shape for all values of $m=0,1,2$ in Fig.~\ref{Fig1}(d). Naturally, for fixed $N$, the energy of the self-trapped modes
increases with vorticity, which drives the modes toward the instability.

\section{QD clusters}

\begin{figure}[b]
\centering
\includegraphics[width=0.45\textwidth]{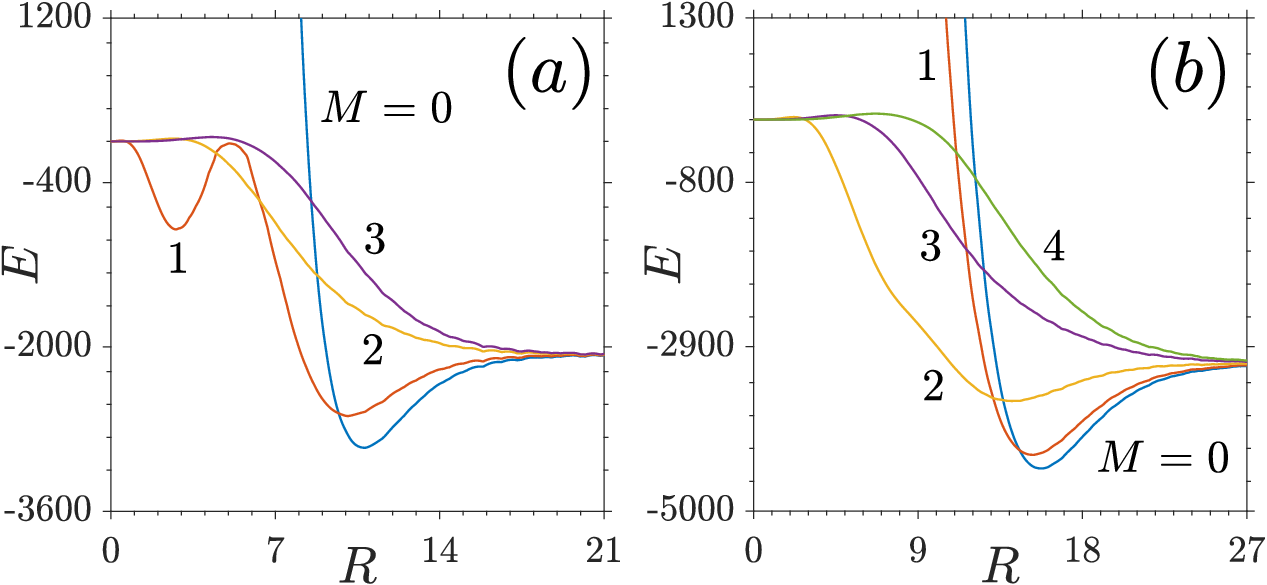}\vskip-0pc
\caption{Energy $E$ of the necklace clusters composed of $\mathcal{N}$ fundamental QDs at $\protect\mu =-0.14$ as a function of the cluster radius $%
R$, calculated as per the input ansatz (\protect\ref{Eq7}) for $\mathcal{N}=6 $ (a) and $\mathcal{N}=9$ (b), for overall charges $M\leq \mathcal{N}/2$.
}
\label{Fig2}
\end{figure}

The droplets shown in Figs.~\ref{Fig1}(a) and (b) are located near the stability boundary for the QDs with $m=1$. These states are optimal for
constructing necklace-shaped clusters composed of QDs [in particular, those corresponding to the blue dots in Fig.~\ref{Fig1}(c)], as the resulting
clusters are metastable and exhibit a minimal radiative loss during the initial stage of the evolution. On the other hand, the radiation becomes
significant when attempting to build clusters from flat-top QDs taken too close to $\mu =\mu _{\text{cut}}$. Further, the clusters built from QDs with
too small values of $|\mu |$ are found to be definitely unstable.

Thus, to construct the necklace clusters, we use the input composed of $\mathcal{N}$ identical fundamental (zero-vorticity) QDs, which are uniformly
placed along the ring with radius $R$:
\begin{equation}
\Psi (t=0)=\mathlarger{\sum}_{n=1}^{\mathcal{N}}w(|\mathbf{r-r_{n}}|)\exp
(2i\pi Mn/\mathcal{N}),  \label{Eq7}
\end{equation}%
where $\mathbf{r_{n}}=\left\{ {R\cos (2\pi n/\mathcal{N}),R\sin (2\pi n/\mathcal{N})}\right\} $ is the center's position of the $n$-th droplet in
plane $z=0$, and $2\pi M/\mathcal{N}$ represents the phase difference between adjacent droplets, where integer $M$ is the overall vorticity of the
cluster. The required phase profile imposed on the droplets constituting the cluster can be realized via phase-imprinting techniques using a far-detuned
broad optical vortex beam \cite{PhysRevLett.89.190403}. The phase in the clusters varies in a steplike fashion along the ring, distinguishing them
from conventional vortex states. In fact, $M$ plays a crucial role in governing the cluster's evolution, as it determines the interaction forces
between adjacent QDs, through the respective phase difference, $\phi =2\pi M/\mathcal{N}$. Specifically, for $M\leq \mathcal{N}/4$ or $M\geq 3\mathcal{N}
/4$ (i.e., $\phi \leq \pi /2$ or $\phi \geq 3\pi /2$), the interaction is attractive \cite{book-2022}, resulting in a net inward force that initially
causes radial contraction of the cluster. For $\mathcal{N}/4<M<3\mathcal{N}/4 $ (corresponding to $\pi /2<\phi <3\pi /2$), the interaction is repulsive,
leading to radial expansion. When $M\neq 0$ and $M\neq N/2$, the cluster possesses nonzero angular momentum
\begin{equation}
J_{z}\approx M\mathcal{N}N  \label{J}
\end{equation}%
[see Eq. (\ref{Eq2}); recall that we consider the symmetric setup, with $N_{1}=N_{2}\equiv N$ being the norm of each component of the individual QD],
hence the radial contraction or expansion is accompanied by rotation. The rotation frequency $\omega $ of a cluster with a large radius $R$ and a
small inner size $\rho $ of each QD, whose inertia moment is $I\approx \mathcal{N}NR^{2}$, can be estimated by equating the corresponding
mechanical angular momentum, $I\omega $, to expression (\ref{J})$,$ which yields
\begin{equation}
\omega \approx M/R^{2}.  \label{om}
\end{equation}%
The comparison of this estimate to numerical findings demonstrates that the agreement is very coarse. For instance, the cluster displayed below
in Fig. \ref{Fig3}(c), which corresponds to $N=6$, $R=10$, and $M=1$, rotates with period $T\approx 400$, while Eq.~(\ref{om}) yields period $2\pi /\omega \simeq 600$. The discrepancy is explained by the fact that the above-mentioned assumption, $\rho \ll R$, does not actually hold for the cluster in Fig. \ref{Fig3}(c).

\begin{figure}[t]
	\centering
	\includegraphics[width=0.45\textwidth,height=0.4\textwidth]{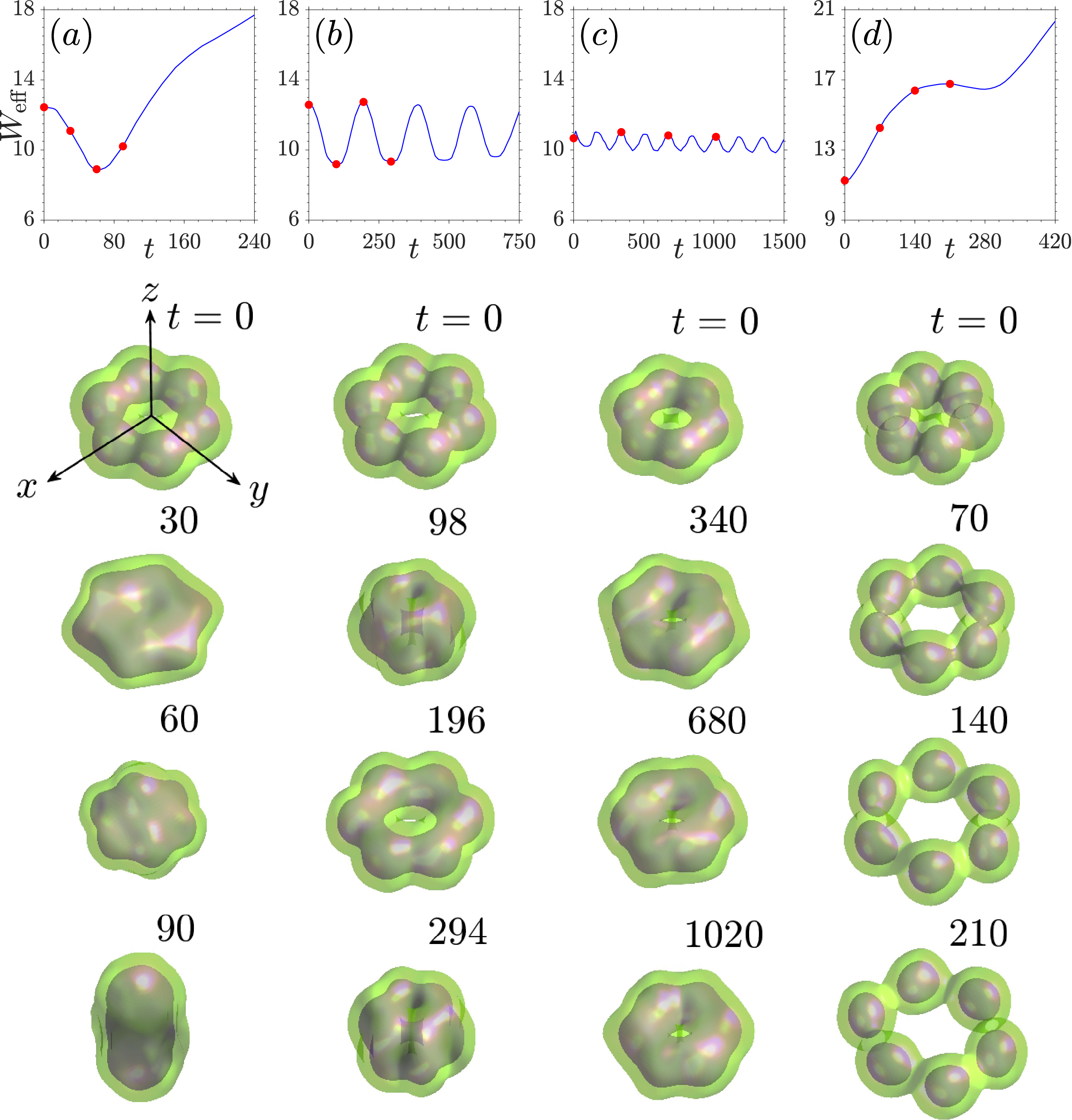}
	\vskip-0pc
	\caption{The evolution of clusters with $\mathcal{N}=6$. (a) The gradual
		fusion into a fundamental state elongated along the $z$-direction at $M=0$, $R=12$. (b) Periodic radial oscillations at $M=1$, $R=12$. The cases
		displayed in panels (a) and (b) correspond to $R>R_{\text{min}}$ [recall $R_{\text{min}}$ is the cluster's radius which provides
		the energy minimum in Fig.~\protect\ref{Fig2}]. (c) The quasi-stationary evolution with the minimal radial variation, provided by setting $R=R_{\text{min}}=10$ at $M=1$. (d) The slow expansion at $M=2$, in the case of $R=10<R_{\text{min}}$. The top row shows the variation of the cluster's radius $W_{\text{eff}}$ [defined as per Eq.~(\protect\ref{W})] as a function of time $t$. The lower rows display isosurface plots of $|\protect\psi |$ at selected times, corresponding to the dots in the top-row plots. Isosurface levels in the 3D figures correspond to $0.15|\protect\psi |_{\mathrm{max}}$ and $0.65|\protect\psi |_{\mathrm{max}}$. }
	\label{Fig3}
\end{figure}

The expected evolution regimes of the clusters can be further inferred from the dependence of energy $E$ of ansatz (\ref{Eq7}) [see Eq. (\ref{Eq2})] on
the initial cluster's radius $R$, as shown in Fig.~\ref{Fig2} for different numbers $\mathcal{N}$ of QDs in the cluster. These dependences indicate
that, for $M\leq \mathcal{N}/4$, an energy minimum occurs at a specific radius $R_{\text{min}}$. The cluster corresponding to the minimum is
energetically favorable and is expected to exhibit minimal shape variations in the course of the evolution, which is corroborated by Fig.~\ref{Fig3}(c). Clusters with $R>R_{\text{min}}$ undergo gradual fusion or exhibit nearly periodic oscillations, as seen in the time dependence of the cluster radius,
\begin{equation}
W_{\text{eff}}=\iint r|w|^{2}dxdy/N,  \label{W}
\end{equation}%
presented in Figs.~\ref{Fig3}(a) and (b). At large values of $R$, for which $E$ remains nearly constant in Fig. \ref{Fig2}, the oscillation period
becomes significantly larger.

For $M=0$, the QDs arranged along the ring with $R>R_{\text{min}}$ tend to move toward the origin, $(r,z)=(0,0)$, and eventually merge, as seen in Fig.~\ref{Fig3}(a). In this case, the effective width $W_{\text{eff}}$ first decreases, attains a minimum, and then increases with time. In the course of
this evolution, the cluster elongates along the $z$-direction [see Fig.~\ref{Fig3}(a)], blurring the initial azimuthal structure. At large $t$, the
increase in $W_{\text{eff}}$ indicates expansion of the cluster in the radial direction. Due to the attractive interactions between adjacent QDs, the clusters with $M=0$ and $R_{\text{min}}$ gradually fuse into a broad fundamental state. The cluster structure cannot be maintained since there is no global orbital angular momentum to counteract the attraction.

For $M=1$, the clusters with $R>R_{\text{min}}$ exhibit radial oscillations accompanied by rotation. As shown by the $W_{\text{eff}}$ curve in Fig.~\ref
{Fig3}(b), the respective oscillation period is $\approx 196$. Although the cluster deforms in the course of the evolution, it maintains the initial
azimuthal structure. Clusters with $R<R_{\text{min}}$ initially expand [see Fig. \ref{Fig3}(d)], and may subsequently exhibit similar
oscillatory behavior.

The most interesting behavior occurs at $R\approx R_{\text{min}}$. In this regime, quasistationary states emerge that undergo persistent rotation with
minimal radial oscillations, as seen in Fig.~\ref{Fig3}(c). The range of $M$ values for which such states can form expands as the number of QDs $\mathcal{%
N}$ in the cluster increases, cf. Figs.~\ref{Fig2}(b) and (a). For clusters with $\mathcal{N}/4<M<3\mathcal{N}/4$, the global energy minimum is reached
only in the limit $R\rightarrow \infty $, which means that such clusters actually expand indefinitely due to the repulsion between adjacent QDs, see
an example in Fig.~\ref{Fig3}(d). Though a local minimal radius $R_{\text{min}}\approx 2.76$ exists for clusters with  $M=1$ [Fig.~\ref{Fig2}(a)], simulation of the evolution reveals that the corresponding cluster is disrupted rapidly due to strong interactions among the closely spaced adjacent QDs.

\begin{figure}[htbp]
\centering
\includegraphics[width=0.45\textwidth,height=0.4\textwidth]{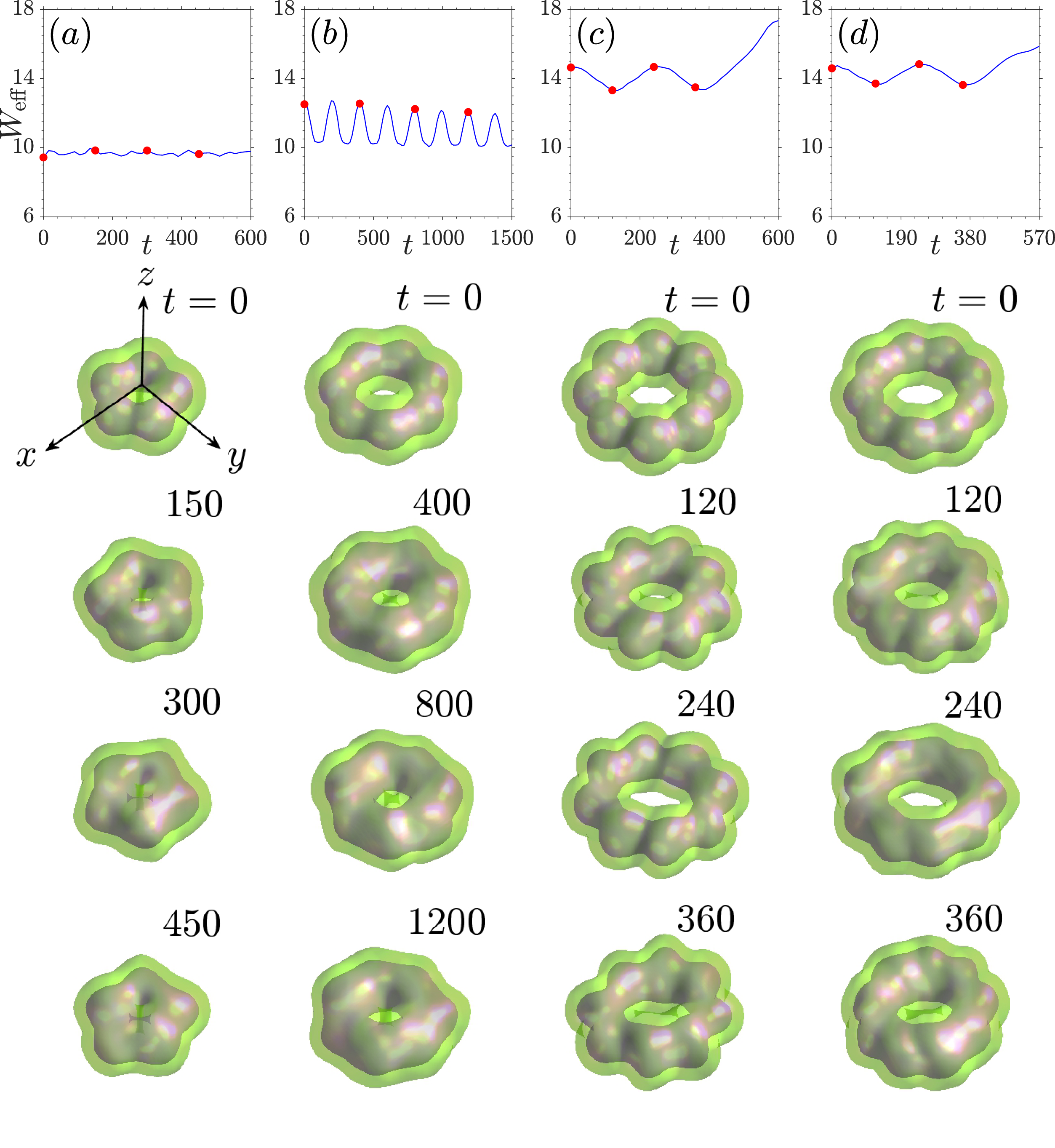}
\caption{(a) The metastable evolution of the cluster with $N=5$, $M=1$, $R=8.5$. (b) Periodic radial oscillations of the cluster with $N=7$, $M=1$, $%
R=12$. (c,d) The robust evolution of the clusters with $M=2$, and $R=14$, composed of $8$ and $9$ droplets, respectively. The top row shows the time
dependence of the cluster's radius $W_{\mathrm{eff}}(t)$, which is defined as per Eq. (\protect\ref{W}). The lower panels display isosurface
plots of $|\protect\psi |$ at selected times, corresponding to the dots in the top-row plots. Isosurface levels in the 3D renderings correspond to $%
0.15|\protect\psi |_{\mathrm{max}}$ and $0.65|\protect\psi |_{\mathrm{max}}$. }
\label{Fig4}
\end{figure}

The central finding of this work is the existence of the robust quasistationary QD clusters (even if they do not exists as genuine
stationary solutions), in comparison to systems with the usual cubic (MF) nonlinearity, where such clusters quickly decay \cite{PhysRevLett.99.133902}. To assess their effective stability in the present system, we added random perturbations (with the amplitude of up to $5\%$) to
the initially constructed clusters. After several cycles of the radial oscillations, the perturbed clusters with $R$ essentially deviating from $R_{\text{min}}$ split into several droplets, which demonstrate a pronounced increase of their width. In contrast, the clusters with initial radii $R\approx R_{\text{min}}$ remain robust against perturbations and persist up to $t>1500$, as demonstrated by Figs.~\ref{Fig3}(c) [this time scale corresponds to $\sim 10$ diffraction (linear-expansion) times, $t_{D}\sim R^{2}$ for the cluster]. Typical metastable clusters are further displayed in Fig.~\ref{Fig4}, with their radii increasing with $\mathcal{N}$. Thus,
the analysis confirms that the minima in the $E(R)$ dependencies in Fig.~\ref{Fig2} correctly predict the radii of the robust necklace clusters, albeit
slightly underestimating the actual radius at which the clusters exhibit the most persistent evolution.

The numerically determined optimal values $R_{\text{min}}$ of radius $R$ were used in all the simulations of the cluster evolution presented in Fig.~%
\ref{Fig4}. The most robust clusters are observed for $\mathcal{N}=5$ [Fig.~\ref{Fig4}(a)] and $\mathcal{N}=6$ [Fig.~\ref{Fig3}(c)]. For other
configurations, the survival time of the perturbed clusters generally decreases with the increase of $\mathcal{N}$. The period of the radial
oscillations also grows with $\mathcal{N}$. All clusters shown in Fig.~\ref{Fig4} exhibit rotation [cf. Eq. (\ref{om})], which is much slower
than the small-amplitude radial oscillations visible in the $W(t)$ dependencies. The rotation period naturally decreases with the increase of $M
$, which imparts the angular momentum to the cluster. Similar findings are expected to be observed in the system including a weak harmonic-oscillator
confinement in the $(x,y)$ plane, which may further expand the variety of families of metastable clusters.

\begin{figure}[tbph]
\centering
\includegraphics[width=0.42\textwidth]{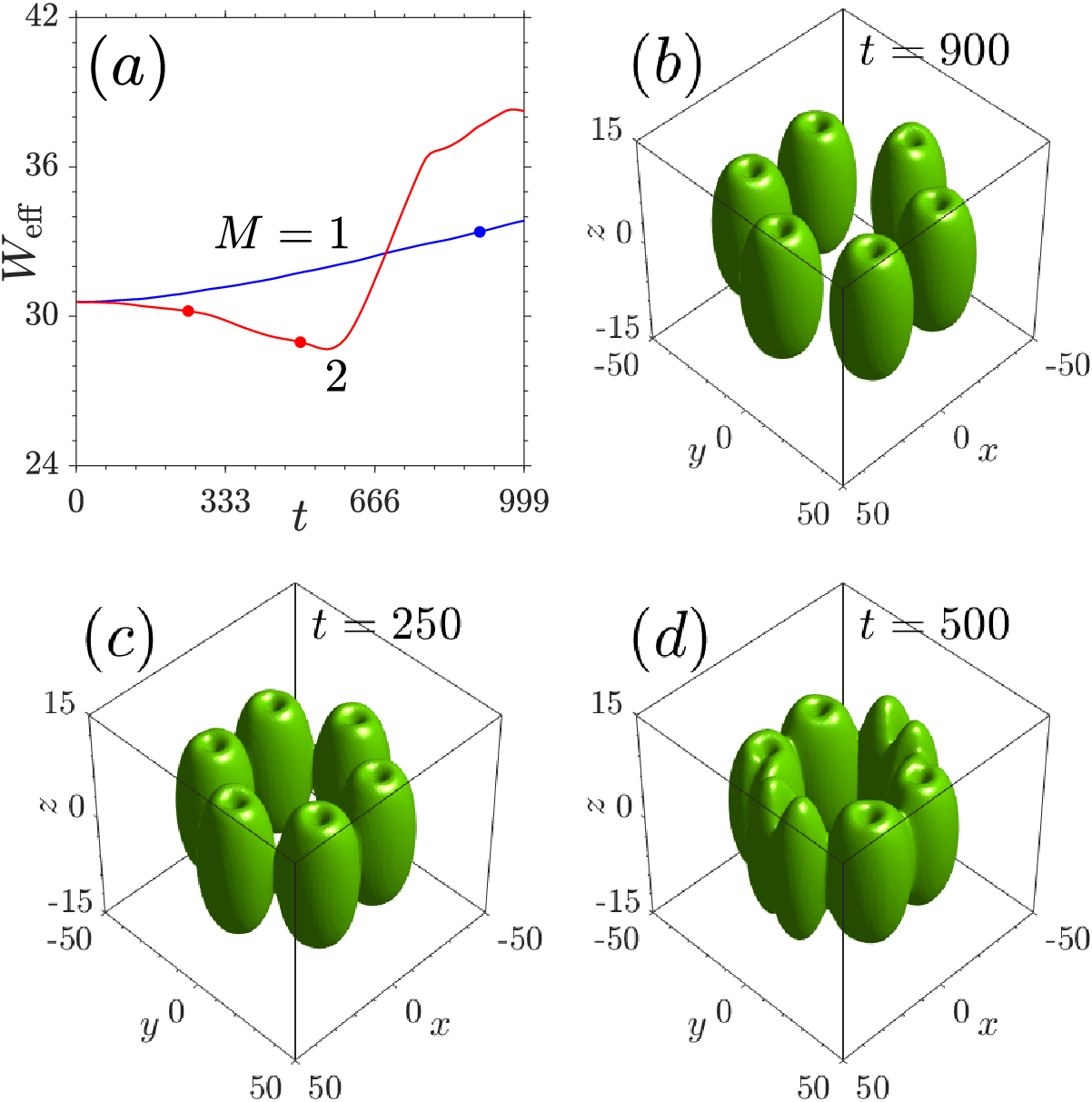}
\caption{The evolution of the \textquotedblleft supervortex" clusters composed of vortex QDs with $m=1$, $\protect\mu =-0.14$, $N=6$, and $R=30$.
(a) The time dependence of the cluster's radius $W_{\mathrm{eff}}(t)$. (b) The slow expansion of the \textquotedblleft supervortex" with $M=1$. (c,d)
The unstable evolution of a \textquotedblleft supervortex" with $M=2$. The isosurfaces in all 3D figures are plotted at $|\protect\psi |=0.05|\protect\psi |_{\mathrm{max}}$ at selected times corresponding to the dots marked in panel (a). }
\label{Fig5}
\end{figure}

To further understand the dynamics of the QD clusters, we construct clusters using the $m=1$ QD from Fig.~\ref{Fig1}(b) as unitary elements. It was
previously demonstrated, in the framework of a 2D model, that such clusters may be built as \textquotedblleft supervortices", with two independent
vorticities: the local one carried by each element ($m=1$, in the present case), and the global vorticity, $M$, which may be imprinted onto the entire
necklace cluster \cite{HS}. Representative examples, including $\mathcal{N}=6$ local-vortex elements, are displayed in Fig.~\ref{Fig5}. The corresponding
dependence of energy $E$ on $R$ for different values of $M$ differs from that presented in Fig.~\ref{Fig2}. No equilibrium cluster radius
corresponding to an energy minimum was found for any value of $M$. At $R\rightarrow \infty $, energy $E$ decreases monotonously, indicating that
the clusters initially composed on a ring with large $R$ may undergo slow expansion, as seen in Fig.~\ref{Fig5}(a). The observed dynamics is affected
by the fact, that, on the contrary to the interaction of two adjacent fundamental solitons (droplets), which is attractive for the phase
difference $\phi =0$ between them and repulsive for $\phi =\pi $, the interaction sign is exactly the opposite for a pair of vortex solitons with
odd values of $m$ \cite{PhysRevE.58.7928}. For $M=1$, such clusters retain their initial structure for a long time [see Fig.~\ref{Fig5}(b)], a behavior
somewhat similar to that of necklace beams formed by $\mathcal{N}$ individual optical solitons \cite{Mihalache_2004}. In contrast, for $M=2$,
the vortex clusters first contract in the radial direction, and then rapidly expand. The latter case exhibits obvious deformation of the constituent
vortex elements and pronounced instability, in Figs.~\ref{Fig5}(c) and (d).

\section{Conclusion}

In summary, we have demonstrated that the fundamental and vortex QDs (quantum droplets) can build ring-shaped clusters in the 3D binary BEC,
which is stabilized by the LHY (Lee-Hung-Yang) correction to the MF (mean-field) contact interaction. The energy-minimization analysis predicts
the equilibrium radius of the clusters. The clusters composed of several fundamental QDs uniformly placed along the ring with the equilibrium radius
stay very robust in the course of the evolution even if the corresponding genuine stationary solution does not exist. When the QDs are
initially placed on a ring whose radius deviates from the equilibrium value, the clusters exhibit contraction, oscillations, or expansion. Imprinting a
global staircase-like phase structure onto the circular clusters slows down the contraction or expansion and induces persistent rotation.
\textquotedblleft Supervortex" clusters, composed of individual QDs with vorticity $m=1$, can also persist for a long time. The QD clusters identified in this study can be generalized for systems with dipole-dipole interactions between magnetic atoms \cite{PhysRevLett.116.215301,NATU2016}, where identical dipolar droplets are uniformly arranged on a ring. Such configurations may exhibit rich dynamics, which should be a subject for a separate work. 

\vskip0.5pc

\vskip0.5pc \textbf{CRediT authorship contribution statement}

Liangwei Dong: Conceptualization and writing the original draft; Dongshuai
Liu: Numerical calculations; Boris A. Malomed: Analytical investigation,
writing, review $\&$ editing and validation.

\vskip0.5pc \textbf{Declaration of competing interest} The authors declare
that they have no known competing financial interests or personal
relationships that could have appeared to influence the work reported in
this paper.

\vskip0.5pc \textbf{Data availability} Data will be made available on
request.

\vskip0.5pc \textbf{Acknowledgment:} This work is supported by the National
Natural Science Foundation of China (NSFC) (grant No. 62575264).


\end{document}